\documentstyle[12pt,epsfig,amssymb]{article}    
\newlength{\dinwidth}                       
\newlength{\dinmargin}                      
\begin{document}
\vspace*{-1.4cm}
\begin{flushright}
DO-TH 96/14\\
RAL-TR-96-061\\
August 1996
\end{flushright}
\renewcommand{\thefootnote}{\fnsymbol{footnote}}
\setcounter{footnote}{1}
%\vspace*{0.4cm}
\begin{center}  \begin{Large} \begin{bf}
Photoproduction of Jets and Heavy Flavors \\
in Polarized {\em{ep}} - Collisions at HERA\footnote{Contribution to the
proceedings of the workshop on 'Future Physics at HERA', DESY,
Hamburg, 1995/96.}\\
  \end{bf}  \end{Large}
  \vspace*{5mm}
  \begin{large}
M.\ Stratmann$^a$, W. Vogelsang$^b$\\
  \end{large}
\end{center}
\renewcommand{\thefootnote}{\arabic{footnote}}
\setcounter{footnote}{0}
$^a$ Universit\"{a}t~Dortmund, Institut~f\"{u}r~Physik, D-44221~Dortmund,
Germany\\
$^b$ Rutherford~Appleton~Laboratory, Chilton, Didcot, Oxon OX11 0QX, England\\
\begin{quotation}
\noindent
{\bf Abstract:}
We study photoproduction of jets and heavy flavors 
in a polarized $ep$ collider mode of HERA at $\sqrt{s}=298$ GeV.
We examine the sensitivity of the cross sections and their asymmetries 
to the proton's polarized gluon distribution and to the completely unknown 
parton distributions of longitudinally polarized photons.
\end{quotation}
\section{Introduction}
HERA has already been very successful in pinning down the proton's
unpolarized gluon distribution $g(x,Q^2)$. Several processes have been 
studied which have contributions from $g(x,Q^2)$ already in the lowest order, 
such as (di)jet and heavy flavor production. Since events at HERA are 
concentrated in the region $Q^2 \rightarrow 0$, the processes have first and 
most accurately been studied in photoproduction [1-6].
As is well-known, in this case the (quasi-real) photon will not only
interact in a direct ('point-like') way, but can also be resolved into 
its hadronic structure. HERA photoproduction experiments like [1-4]
have not merely established evidence for the existence of such a resolved
contribution, but have also been precise enough to improve our knowledge
about the parton distributions, $f^{\gamma}$, of the photon.

Given the success of such unpolarized photoproduction experiments at HERA, 
it seems most promising \cite{wir} to closely examine the same processes for 
the situation with longitudinally polarized beams with regard to their 
sensitivity to the proton's polarized gluon distribution $\Delta g$, which is
still one of the most interesting, but least known, quantities in
'spin-physics'. Recent next-to-leading (NLO) studies of polarized
DIS \cite{grsv,gs} show that the $x$-shape of $\Delta g$ seems to be 
hardly constrained at all by the present DIS data \cite{data}, even though
a tendency towards a sizeable positive {\em total} gluon polarization,
$\int_0^1 \Delta g(x,Q^2=4 \; \mbox{GeV}^2) dx \gtrsim 1$, was found
\cite{grsv,bfr,gs}. Furthermore, polarized photoproduction experiments 
may in principle allow to not only determine the parton, in
particular gluon, content of the polarized {\em proton}, but also 
that of the longitudinally polarized {\em photon} which is completely
unknown so far. Since, e.g., a measurement of the photon's spin-dependent
structure function $g_1^{\gamma}$ in polarized $e^+ e^-$ collisions is not
planned in the near future, HERA could play a unique role here, even if it
should only succeed in establishing the very {\em existence} of a resolved
contribution to polarized photon-proton reactions.

Our contribution is organized as follows: In section 2 we collect the 
necessary ingredients for our calculations. Section 3 is devoted to open-charm
photoproduction. In section 4 we examine polarized photoproduction of 
(di)jets. Section 5 contains the conclusions.
%
% PDFS 
%
\section{Polarized Parton Distributions of the Proton and the Photon}
Even though NLO analyses of polarized DIS which take into account all or 
most data sets \cite{data} have been published recently \cite{grsv,bfr,gs},
we have to stick to LO calculations throughout this work since the NLO 
corrections to polarized charm or jet production are not yet known. This 
implies use of LO parton distributions, which have also been provided in the 
studies \cite{grsv,gs}. Both papers give various LO sets which mainly differ 
in the $x$-shape of the polarized gluon distribution. We will choose the LO 
'valence' set of the 'radiative parton model analysis' \cite{grsv}, which 
corresponds to the best-fit result of that paper, along with two other sets 
of \cite{grsv} which are based on either assuming $\Delta g (x,\mu^2) = 
g(x,\mu^2)$ or $\Delta g(x,\mu^2)=0$ at the low input scale $\mu$ of 
\cite{grsv}, where $g(x,\mu^2)$ is the unpolarized LO GRV \cite{grv} input 
gluon distribution. These two sets will be called '$\Delta g=g$ input' and 
'$\Delta g=0$ input' scenarios, respectively. The gluon of set C of \cite{gs}
is qualitatively different since it has a substantial negative polarization 
at large $x$. We will therefore also use this set in our calculations.
For illustration, we show in Fig.~1 the gluon distributions  
of the four different sets of parton distributions we will use, taking a 
typical scale $Q^2=10$ GeV$^2$. Keeping in mind that all four LO sets provide 
very good descriptions of all present polarized DIS data \cite{data}, 
it becomes obvious that the data indeed do not seem to be able to 
significantly constrain the $x$-shape of $\Delta g(x,Q^2)$.
\begin{figure}[ht]
\begin{center}
\vspace*{-1.5cm}
\epsfig{file=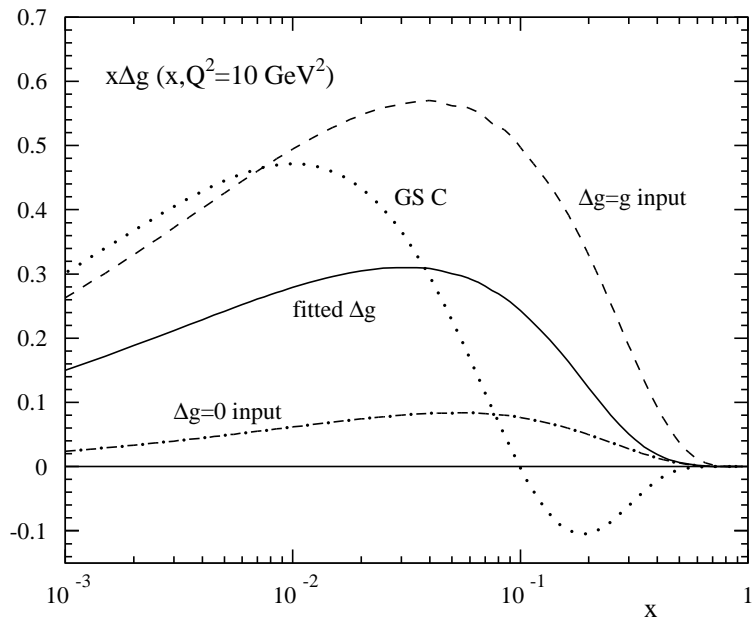,width=11.5cm}
\vspace*{-1.3cm}
\caption{\it Gluon distributions at $Q^2=10\,{\it{GeV}}^{\:2}$ 
of the four LO sets of
polarized parton distributions used in this paper. The dotted line refers
to set C of \cite{gs}, whereas the other distributions are taken 
from \cite{grsv} as described in the text.}
\vspace*{-0.5cm}
\end{center}
\end{figure}

In the case of photoproduction the electron just
serves as a source of quasi-real photons which are radiated according 
to the Weizs\"{a}cker-Williams spectrum \cite{ww}. The photons can then 
interact either directly or via their partonic structure ('resolved' 
contribution). In the case of longitudinally polarized electrons, the 
resulting photon will be longitudinally (more precisely, circularly)
polarized and, in the resolved case, the {\em polarized} parton 
distributions of the photon, $\Delta f^{\gamma}(x,Q^2)$, enter the 
calculations. Thus one can define the effective polarized parton densities 
at the scale $M$ in the longitudinally polarized electron via\footnote{We 
include here the additional definition 
$\Delta f^{\gamma} (x_{\gamma},M^2) \equiv \delta (1-x_{\gamma})$ for the 
direct ('unresolved') case.}
\begin{equation}  \label{elec}
\Delta f^e (x_e,M^2) = \int_{x_e}^1 \frac{dy}{y} \Delta P_{\gamma/e} (y)
\Delta f^{\gamma} (x_{\gamma}=\frac{x_e}{y},M^2) \; 
\end{equation}
($f=q,g$) where $\Delta P_{\gamma/e}$ is the polarized Weizs\"{a}cker-Williams
spectrum for which we will use 
\begin{equation}  \label{weiz}
\Delta P_{\gamma/e} (y) = \frac{\alpha_{em}}{2\pi} \left[ 
\frac{1-(1-y)^2}{y} \right] \ln \frac{Q^2_{max} (1-y)}{m_e^2 y^2} \; ,
\end{equation}
with the electron mass $m_e$. For the time being, it seems most 
sensible to follow as closely as possible the analyses successfully 
performed in the unpolarized case, which implies to introduce the same 
kinematical cuts. As in \cite{jet1ph,jet2ph,chph,kramer} we will use
an upper cut\footnote{In H1 analyses of HERA photoproduction
data \cite{jet1h1,jet2h1,chh1}
the cut $Q^2_{max}=0.01$ GeV$^2$ is used along with slightly
different $y$-cuts as compared to the corresponding 
ZEUS measurements \cite{jet1ph,jet2ph,chph}, which leads to
smaller rates.} 
$Q^2_{max}=4$ GeV$^2$, and the $y$-cuts $0.2 \leq y \leq 0.85$ 
(for charm and single-jet \cite{jet1ph} production) 
and $0.2 \leq y \leq 0.8$ (for dijet production, \cite{jet2ph}) 
will be imposed. We note that a larger value for the lower limit, $y_{min}$, 
of the allowed $y$-interval would enhance the yield of polarized photons 
relative to that of unpolarized ones since $\Delta P_{\gamma/e}(y)/
P_{\gamma/e}(y)$, where $P_{\gamma/e}$ is the unpolarized 
Weizs\"{a}cker-Williams spectrum obtained by using $[ (1+(1-y)^2)/y ]$
instead of the square bracket in (\ref{weiz}),
is suppressed for small $y$. On the other hand, increasing $y_{min}$ 
would be at the expense of reducing the individual polarized and
unpolarized rates.

The polarized photon structure functions $\Delta f^{\gamma} (x_{\gamma},M^2)$ 
in (\ref{elec}) are completely unmeasured so far, such that 
models for them have to be invoked. To obtain a realistic estimate for the 
theoretical uncertainties in the polarized photonic parton densities
two very different scenarios were considered in \cite{gvg,gsvg} assuming 
'maximal' ($\Delta f^{\gamma}(x,\mu^2)=f^{\gamma}(x,\mu^2)$) or 'minimal' 
($\Delta f^{\gamma}(x,\mu^2)=0$) saturation of the fundamental positivity 
constraints $|\Delta f^{\gamma}(x,\mu^2)| \leq f^{\gamma}(x,\mu^2)$ at the
input scale $\mu$ for the QCD evolution. Here $\mu$ and the unpolarized 
photon structure functions $f^{\gamma}(x,\mu^2)$ were adopted from the 
phenomenologically successful radiative parton model predictions in 
\cite{grvg}. The results of these two extreme approaches are presented in 
Fig.~2
\begin{figure}[hbt]
\begin{center}
\vspace*{-1.5cm}
\epsfig{file=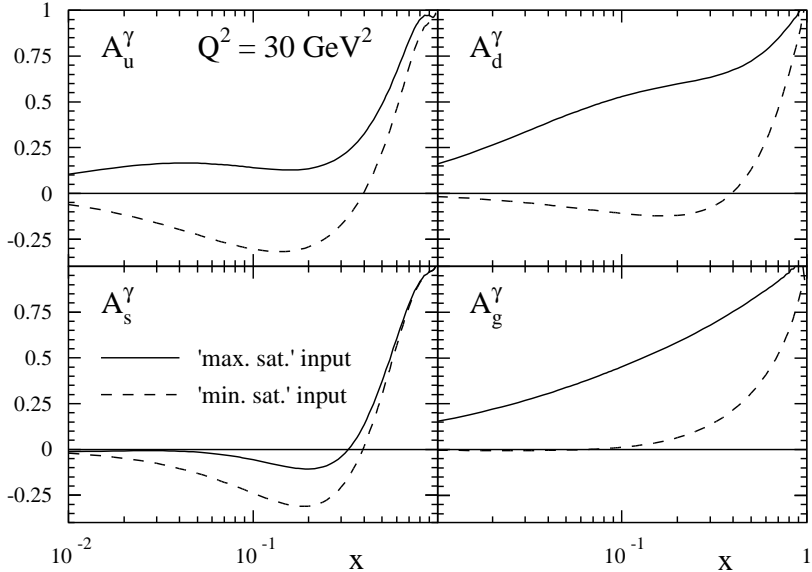,width=12cm}
\vspace*{-1cm}
\caption{\it Photonic LO parton asymmetries 
$A_f^{\gamma}\equiv \Delta f^{\gamma}/f^{\gamma}$ at 
$Q^2=30\,{\it{GeV}}^{\,2}$ 
for the two scenarios considered in \cite{gvg,gsvg} (see text). The 
unpolarized LO photonic parton distributions were taken from \cite{grvg}.}
\vspace*{-0.5cm}
\end{center}
\end{figure}
in terms of the photonic parton asymmetries $A_f^{\gamma} \equiv 
\Delta f^{\gamma}/f^{\gamma}$, evolved to $Q^2=30$ GeV$^2$ in LO. An ideal 
aim of measurements in a polarized collider mode of HERA would of course be 
to determine the $\Delta f^{\gamma}$ and to see which ansatz is more 
realistic. The sets presented in Fig.~2, which we 
will use in what follows, should in any case be sufficient to study 
the sensitivity of the various cross sections to the $\Delta f^{\gamma}$,
but also to see in how far they influence a determination of $\Delta g$.

We finally note that in what follows a polarized cross section will always
be defined as
\begin{equation} 
\Delta \sigma \equiv \frac{1}{2} \left( \sigma (++)-\sigma (+-) \right) \; ,
\end{equation}
the signs denoting the helicities of the scattering particles. The
corresponding unpolarized cross section is given by taking the sum 
instead, and the cross section asymmetry is $A\equiv \Delta \sigma/\sigma$.
Whenever calculating an asymmetry $A$, we will 
use the LO GRV parton distributions for the proton \cite{grv} and the 
photon \cite{grvg} to calculate the unpolarized cross section.
For consistency, we will employ the LO expression for the strong coupling 
$\alpha_s$ with \cite{grsv,gs,gvg,gsvg} $\Lambda_{QCD}^{(f=4)}=200$ MeV for 
four active flavors.
%
%charm
%
\section{Charm Photoproduction at HERA}
For illustration, we first briefly consider the total cross section. In the 
unpolarized case it has been possible to extract the total cross section for 
$\gamma p \rightarrow c\bar{c}$ from the fixed target \cite{oldch} and HERA 
\cite{chph,chh1} lepton-nucleon data, i.e., the open-charm cross section for a 
fixed photon energy without the smearing from the Weizs\"{a}cker-Williams 
spectrum. To LO, the corresponding polarized cross section is given by 
\begin{equation}   \label{wqctot}
\Delta \sigma^c (S_{\gamma p}) =\sum_{f^{\gamma},f^p} 
\int_{4m_c^2/S_{\gamma p}}^1
d x_{\gamma} \int_{4m_c^2/x_{\gamma}S_{\gamma p}}^1 d x_p
\Delta f^{\gamma} (x_{\gamma},M^2) \Delta f^p (x_p,M^2)
\Delta \hat{\sigma}^c (\hat{s},M^2) \; .
\end{equation}
where $M$ is some mass scale and $\hat{s} \equiv x_{\gamma} 
x_p S_{\gamma p}$. The $\Delta f^p$ stand for the polarized parton 
distributions of the proton. In the direct case, the contributing subprocess 
is photon-gluon fusion (PGF), $\gamma g \rightarrow c\bar{c}$, whose 
spin-dependent total LO subprocess cross section $\Delta \hat{\sigma}^c 
(\hat{s})$ can be found in \cite{gr,grvalt}. In the resolved case, the 
processes $gg \rightarrow c\bar{c}$ and $q\bar{q} 
\rightarrow c\bar{c}$ contribute; their cross sections have been calculated 
in \cite{cont}. Needless to say that we can obtain the corresponding 
unpolarized LO charm cross section $\sigma^c (S_{\gamma p})$ by using 
LO unpolarized parton distributions and subprocess cross sections 
(as calculated in \cite{chunp}) in (\ref{wqctot}). 

Tab.~1 shows our results \cite{wir} for the 
polarized cross sections and the asymmetries $\Delta \sigma^c/\sigma^c$ for 
the four different sets of polarized parton distributions, where we have
used the scale $M=2 m_c$ with the charm mass $m_c=1.5$ GeV. The resolved 
contribution to the cross section is rather small in the unpolarized case. 
For the polarized case, we have calculated it using the 'maximally' 
saturated set for the polarized photon structure functions, which should 
roughly provide the maximally possible background from resolved photons. 
The corresponding results are shown individually in Tab.~1. The resolved 
contribution turns out to be non-negligible only for large 
$\sqrt{S_{\gamma p}}$, where it can be as large as about 1/3 the direct 
contribution but with opposite sign. As becomes obvious from Tab.~1 (see also 
\cite{fr} and Fig.~4 of \cite{wir}), the asymmetry becomes very small towards 
the HERA region at larger $\sqrt{S_{\gamma p}} \sim 200$ GeV. One reason for 
this is the oscillation of the polarized subprocess cross section for the 
direct part, combined with cancellations between the direct and the resolved 
parts. More importantly, as seen from (\ref{wqctot}), the larger 
$S_{\gamma p}$ becomes, the smaller are the $x_{p,\gamma}$ - values probed, 
such that the rapid rise of the unpolarized parton distributions strongly 
suppresses the asymmetry. The smallness of the asymmetries and the possibly 
significant influence of the resolved contribution on them will make their 
measurement and a distinction between the different $\Delta g$ 
elusive\footnote{As was shown in \cite{fr}, the situation at HERA energies 
somewhat improves if cuts on the produced charm quark's transverse momentum 
and rapidity are imposed. However, the polarized resolved contribution to the 
asymmetry was neglected in \cite{fr}.}. The measurement of the {\em total} 
charm cross section asymmetry in $\gamma p \rightarrow c\bar{c}$ seems rather 
more feasible at smaller energies, $\sqrt{S_{\gamma p}} \lesssim 20$ GeV, 
i.e., in the region accessible by the future COMPASS experiment \cite{compass}
where also the unknown resolved contribution to the cross section is 
negligible. 
\noindent
\begin{table}[htb]
%%%%%%%%%%%%%%%%%%%%%%%%%%%
%        TABLE
%%%%%%%%%%%%%%%%%%%%%%%%%%%
%
\hbox to \textwidth {\hss
\begin{tabular}{c|c|c|c||c|c|c||c|c|c||c|c|c|}
&\multicolumn{3}{c||}{fitted $\Delta g$}&
 \multicolumn{3}{c||}{$\Delta g = g$ input} &
 \multicolumn{3}{c||}{$\Delta g = 0$ input} &
 \multicolumn{3}{c|}{GS C} \\ \cline{2-13} 
$\sqrt{S_{\gamma p}}$&
\multicolumn{1}{c}{dir.} &
\multicolumn{1}{c}{res.} &
\multicolumn{1}{c||}{$A$} &
\multicolumn{1}{c}{dir.} &
\multicolumn{1}{c}{res.} &  
\multicolumn{1}{c||}{$A$} &
\multicolumn{1}{c}{dir.} &
\multicolumn{1}{c}{res.} &
\multicolumn{1}{c||}{$A$} &
\multicolumn{1}{c}{dir.} &
\multicolumn{1}{c}{res.} & 
\multicolumn{1}{c|}{$A$} \\
$\mbox{[GeV]}$& 
\multicolumn{1}{c}{$\mbox{[nb]}$} &
\multicolumn{1}{c}{$\mbox{[nb]}$} &
\multicolumn{1}{c||} {$[\%]$}& 
\multicolumn{1}{c}{$\mbox{[nb]}$} &
\multicolumn{1}{c}{$\mbox{[nb]}$} &  
\multicolumn{1}{c||} {$[\%]$}&
\multicolumn{1}{c}{$\mbox{[nb]}$} &
\multicolumn{1}{c}{$\mbox{[nb]}$} &
\multicolumn{1}{c||} {$[\%]$}&
\multicolumn{1}{c}{$\mbox{[nb]}$} &
\multicolumn{1}{c}{$\mbox{[nb]}$} &
\multicolumn{1}{c|} {$[\%]$}
\\
\hline
\hline
20&13.9&-0.29 & 2.6 & 23.2&0.33& 4.5 &3.26 & -0.56 & 0.52& 19.6 &
-0.70& 3.6 \\ \hline
50&2.07&1.30 & 0.18 &1.15&2.94& 0.21 &-0.53&0.18& -0.019 &
17.4&0.61& 0.94\\ \hline
200&-7.00&2.06& -0.06&-12.8&3.60&-0.11 &-1.96&0.44& -0.018&-8.79&3.23&
-0.067 \\ \hline
\end{tabular}
\hss}
\caption{\it Total cross sections and 
asymmetries $A$ for charm photoproduction in polarized $\gamma p$ collisions.}
\end{table}

\hspace*{0cm}From our observations for HERA-energies it follows that it 
could be more promising to study distributions of the cross section in the 
transverse momentum or the rapidity of the charm quark in order to 
cut out the contributions from very small $x_{p,\gamma}$. We will now 
include the Weizs\"{a}cker-Williams spectrum since tagging of the electron, 
needed for the extraction of the cross section at fixed photon energy, will 
probably reduce the cross section too strongly. The polarized LO cross 
section for producing a charm quark with transverse momentum $p_T$ and 
cms-rapidity $\eta$ then reads 
\begin{equation} \label{wqc}
\frac{d^2 \Delta \sigma^c}{dp_T d\eta} = 2 p_T 
\sum_{f^e,f^p} \int_{\frac{\rho e^{-\eta}}
{1-\rho e^{\eta}}}^1 d x_e x_e \Delta f^e (x_e,M^2) x_p \Delta f^p (x_p,M^2)
\frac{1}{x_e - \rho e^{-\eta}} \frac{d\Delta \hat{\sigma}}{d\hat{t}} \; ,
\end{equation}
where $\rho \equiv m_T/\sqrt{S}$ with $m_T \equiv \sqrt{p_T^2+m_c^2}$, and 
$x_p \equiv x_e \rho e^{\eta}/(x_e - \rho e^{-\eta})$.
The cross section can be transformed to the more relevant HERA
laboratory frame by a simple boost which implies 
$\eta \equiv \eta_{cms} = \eta_{LAB} -\frac{1}{2} \ln (E_p/E_e)$,
where we have, as usual, counted positive rapidity in the proton
forward direction. The spin-dependent differential LO subprocess cross 
sections $d\Delta \hat{\sigma}/d\hat{t}$ for the resolved processes 
$gg \rightarrow c\bar{c}$ and $q\bar{q} \rightarrow c\bar{c}$ with $m_c 
\neq 0$ can again be found in \cite{cont}. For the
factorization/renormalization scale in (\ref{wqc}) we choose $M=m_T/2$; 
we will comment on the scale dependence of the results at the end of this 
section.

\begin{figure}[ht]
\begin{center}
\vspace*{-1.5cm}
\epsfig{file=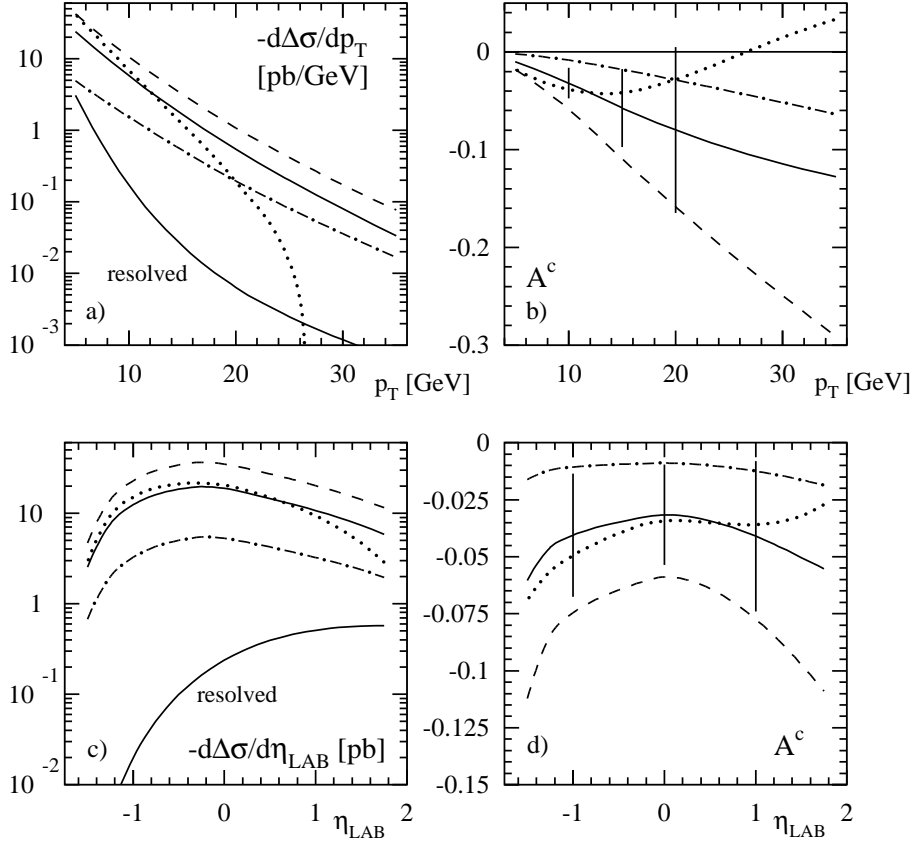,width=13.cm}
\vspace*{-0.7cm}
\caption{\it {\bf a:} $p_T$-dependence of the (negative) polarized 
charm-photoproduction 
cross section in $ep$-collisions at HERA, calculated according to 
(\ref{wqc}) (using $M=m_T/2$ and $m_c=1.5$ GeV) and integrated over 
$-1 \leq \eta_{LAB} \leq 2$. The line drawings are as in Fig.~1.
For comparison the resolved contribution to the cross section, calculated
with the 'fitted $\Delta g$' gluon distribution of \cite{grsv} and the 
'maximally' saturated set of polarized photonic parton distributions is
shown by the lower solid line. {\bf b:} Asymmetry corresponding to {\bf a}.
The expected statistical errors indicated by the bars have been calculated
according to (\ref{aerr}) and as explained in the text.
{\bf c,d:} Same as {\bf a,b}, but for the $\eta_{LAB}$-dependence, integrated
over $p_T>8$ GeV.}
\vspace*{-0.5cm}
\end{center}
\end{figure}
Fig.~3 shows our results \cite{wir} obtained for the four different sets of 
polarized parton distributions for $E_p=820$ GeV and $E_e=27$ GeV.  
Fig.~3a displays the $p_T$-dependence of the cross section, where we have 
integrated over $-1\leq \eta_{LAB} \leq 2$. The resolved contribution to the 
cross section has been included, calculated with the 'maximally' saturated 
set of polarized photon structure functions. It is shown individually for
the 'fitted $\Delta g$'-set of polarized proton distributions by the lower 
solid line in Fig.~3a.
Comparison of the two solid lines in Fig.~3a shows that the resolved
contribution is negligibly small in this case unless $p_T$ becomes very
small. Fig.~3b shows the asymmetries corresponding to Fig.~3a. It becomes 
obvious that they are much larger than for the total cross section 
if one goes to $p_T$ of about 10-20 GeV, which is in agreement 
with the corresponding findings of \cite{fr}. Furthermore, one sees that 
the asymmetries are strongly sensitive to the size {\em and} shape of the 
polarized gluon distribution used. Similar statements are true for the 
$\eta_{LAB}$-distributions shown in Figs.~3c,d, where $p_T$ has been 
integrated over $p_T>8$ GeV in order to increase the number of events. 
Even here the resolved contribution remains small, although it becomes 
more important towards large positive values of $\eta_{LAB}$. 
We have included in the asymmetry plots in Figs.~3b,d the expected 
statistical errors $\delta A$ at HERA which can be estimated from 
\begin{equation}  \label{aerr}
\delta A = \frac{1}{P_e P_p \sqrt{{\cal L} \sigma \epsilon}} \; ,
\end{equation}
where $P_e$, $P_p$ are the beam polarizations, ${\cal L}$ is the integrated 
luminosity and $\epsilon$ the charm detection efficiency, for which we assume
$P_e * P_p=0.5$, ${\cal L}=100$/pb and $\epsilon=0.15$. The error
bars in Fig.~3b,d have been obtained by integrating the unpolarized LO
cross sections $d\sigma/dp_T$ or $d\sigma/d\eta_{LAB}$ over bins of 
$\Delta p_T=5$ GeV or $\Delta \eta_{LAB}=1$, respectively, and have 
been plotted at the centers of the bins. It becomes obvious that it 
will be quite difficult to distinguish between different gluon distributions
in our proposed charm experiments. The situation would, however, become
much better for a higher luminosity of, say, ${\cal L}=1000$/pb in which 
case the error bars would decrease by a factor 3.

We now briefly address the theoretical uncertainties of our results 
in Fig.~3 related to the dependence of the cross sections and asymmetries 
on the renormalization/factorization scale $M$. Since all our calculations
could be performed in LO only, this is a particularly important issue. 
When using the scale $M=m_T$ it turns out that the cross 
sections in Fig.~3a are subject to changes of about $10\%$ at $p_T<15$ GeV,
and of as much as $20-25\%$ at larger $p_T$. Changes of in most cases below
$10\%$ are found for the $\eta_{LAB}$-curves in Fig.~3c. In contrast to this 
(not unexpected) fairly strong scale dependence of the polarized cross 
sections, the {\em asymmetries}, 
which will be the quantities actually measured, are very insensitive to 
scale changes, deviating usually by not more than a few percent from the 
values shown in Fig.~3b,d for all relevant $p_T$ and  $\eta_{LAB}$. 
This finding seems important in two respects: Firstly, 
it warrants the genuine sensitivity of the asymmetry to $\Delta g$, implying
that despite the sizeable scale dependence of the cross section it still seems
a reasonable and safe procedure to compare LO theoretical predictions for the 
asymmetry with future data and to extract $\Delta g$ from such a comparison.
Secondly, it sheds light on the possible role of NLO corrections to 
our results, suggesting that such corrections might be sizeable for the 
cross sections, but less important for the asymmetry. 
%
%jets
%
\section{Photoproduction of Jets}
The generic cross section formula for the production of a single jet 
with transverse momentum $p_T$ and rapidity $\eta$ is similar to
that in (\ref{wqc}), the sum now running over all properly symmetrized 
$2\rightarrow 2$ subprocesses for the direct ($\gamma b\rightarrow cd$) 
and resolved ($ab\rightarrow cd$) cases. When only light flavors are 
involved, the corresponding differential helicity-dependent LO subprocess 
cross sections can be found in \cite{bab}. In all following predictions 
we will deal with the charm contribution to the cross section by including 
charm only as a {\em{final}} state particle produced via the subprocesses 
$\gamma g \rightarrow c\bar{c}$ (for the direct part) and $gg \rightarrow 
c\bar{c}$, $q\bar{q} \rightarrow c\bar{c}$ (for the resolved part). For the 
values of $p_T$ considered it turns out that the finite charm mass can be 
safely neglected. In all following applications we will use the 
renormalization/factorization scale $M=p_T$. We have again found that the 
scale dependence of the asymmetries is rather weak as compared to that of the 
cross sections.

It appears very promising \cite{wir} to study the $\eta_{LAB}$-distribution of 
the cross section and the asymmetry. The reason for this is that for negative 
$\eta_{LAB}$ the main contributions are expected to come from the region
$x_{\gamma} \rightarrow 1$ and thus mostly from the direct piece
at $x_{\gamma}=1$. To investigate this, Fig.~4 shows our results for 
\begin{figure}[ht]
\begin{center}
\vspace*{-1.5cm}
\epsfig{file=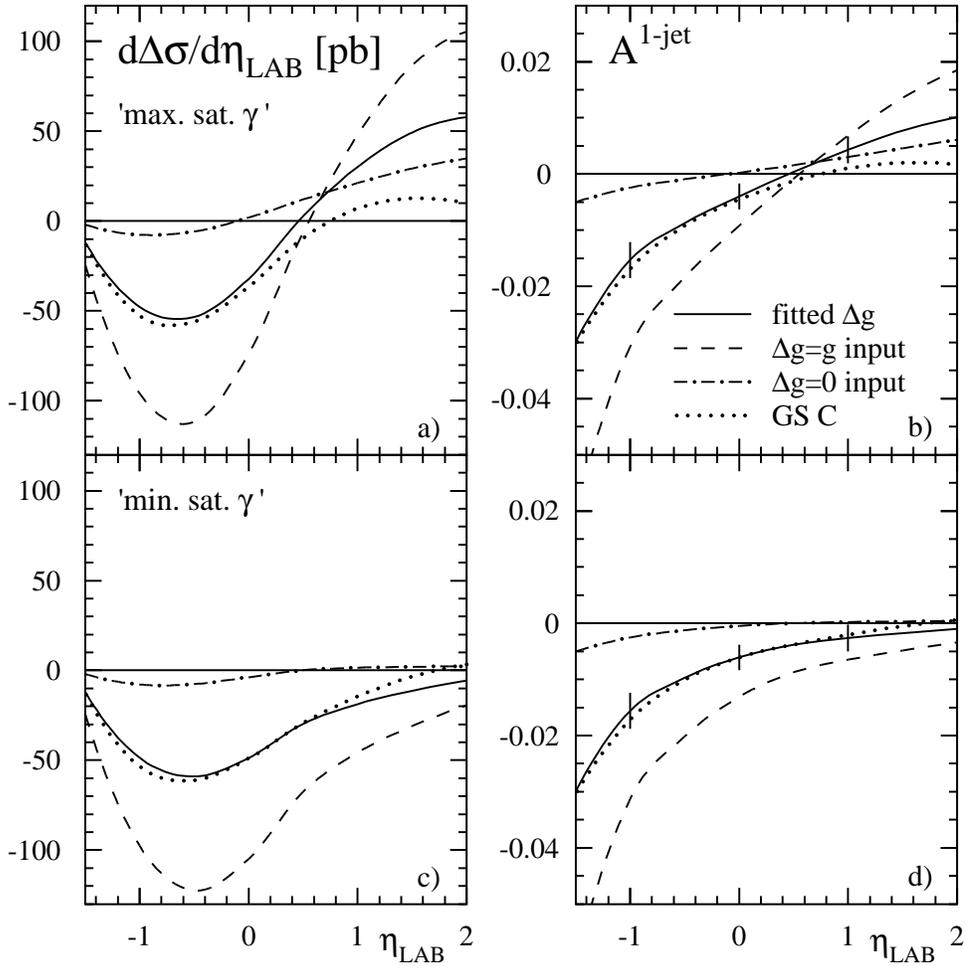,width=14.5cm}
\vspace*{-1.0cm}
\caption{\it {\bf a:} $\eta_{LAB}$-dependence of the polarized single-jet 
inclusive photoproduction cross section in $ep$-collisions at HERA, integrated
over $p_T > 6$ GeV. The renormalization/factorization scale was 
chosen to be $M=p_T$. The resolved contribution to the cross section has 
been calculated with the 'maximally' saturated set of polarized photonic 
parton distributions. {\bf b:} Asymmetry corresponding to {\bf a}. The 
expected statistical errors have been calculated according to (\ref{aerr})
and as described in the text. {\bf c,d:} Same as {\bf a,b}, 
but for the 'minimally' saturated set of polarized photonic parton 
distributions.}
\vspace*{-0.5cm}
\end{center}
\end{figure}
the single-inclusive jet cross section and its asymmetry vs. $\eta_{LAB}$ 
and integrated over $p_T>8$ GeV for the four sets of the polarized 
proton's parton distributions. For Figs.~4a,b we have used the 'maximally' 
saturated set of polarized photonic parton densities, whereas Figs.~4c,d 
correspond to the 'minimally' saturated one. Comparison of Figs.~4a,c or
4b,d shows that indeed the direct contribution clearly dominates for 
$\eta_{LAB} \leq -0.5$, where also differences between the polarized gluon 
distributions of the proton show up clearly. Furthermore, the cross sections 
are generally large in this region with asymmetries of a few percents. At 
positive $\eta_{LAB}$, we find that the cross section is 
dominated by the resolved contribution and is therefore sensitive to
both the parton content of the polarized proton {\em and} the photon.
This means that one can only learn something about the polarized photon 
structure functions if the polarized parton distributions of the proton are 
already known to some accuracy or if an experimental distinction between 
resolved and direct contributions can be achieved. We note that the dominant 
contributions to the resolved part at large $\eta_{LAB}$ are driven by the 
polarized photonic {\em gluon} distribution $\Delta g^{\gamma}$. Again we 
include in Figs.~4b,d the expected statistical errors which we have estimated 
according to (\ref{aerr}) with $P_e * P_p=0.5$, ${\cal L}=100$/pb, 
$\epsilon=1$ for bins of $\Delta \eta_{LAB}=1$. From the results it appears 
that a measurement of the proton's $\Delta g$ should be possible from 
single-jet events at negative rapidities where the contamination from the 
resolved contribution is minimal. 

In the unpolarized case, an experimental criterion for a distinction  
between direct and resolved contributions has been introduced \cite{jeff} and 
used \cite{jet2ph} in the case of dijet photoproduction at HERA. We will now 
adopt this criterion for the polarized case to see whether it would enable a
further access to $\Delta g$ and/or the polarized photon structure functions. 
The generic expression for the polarized cross section for the photoproduction
of two jets with laboratory system rapidities $\eta_1$, $\eta_2$ is to LO
\begin{equation} \label{wq2jet}
\frac{d^3 \Delta \sigma}{dp_T d\eta_1 d\eta_2} = 2 p_T 
\sum_{f^e,f^p} x_e \Delta f^e (x_e,M^2) x_p \Delta f^p (x_p,M^2)
\frac{d\Delta \hat{\sigma}}{d\hat{t}} \; ,
\end{equation}
where $p_T$ is the transverse momentum of one of the two jets (which balance
each other in LO) and 
\begin{equation}
x_e \equiv \frac{p_T}{2 E_e} \left( e^{-\eta_1} + e^{-\eta_2} \right)\;\; , \;
x_p \equiv \frac{p_T}{2 E_p} \left( e^{\eta_1} + e^{\eta_2} \right) \; .
\end{equation}
Following \cite{jet2ph}, we will integrate over the cross section to obtain 
$d\Delta \sigma/d\bar{\eta}$, where $\bar{\eta} \equiv (\eta_1 + \eta_2)/2$.
Furthermore, we will apply the cuts \cite{jet2ph}
$|\Delta \eta| \equiv |\eta_1-\eta_2| \leq 0.5 \; , \;\; 
p_T>6 \; \mbox{GeV}$.
The important point is that measurement of the jet rapidities allows 
for fully reconstructing the kinematics of the underlying hard subprocess
and thus for determining the variable \cite{jet2ph}
\begin{equation}
x_{\gamma}^{OBS} = \frac{\sum_{jets} p_T^{jet} e^{-\eta^{jet}}}{2yE_e} \; ,
\end{equation} 
which in LO equals $x_{\gamma}=x_e/y$ with $y$ as before being the 
fraction of the electron's energy taken by the photon. Thus it becomes
possible to experimentally select events at large $x_{\gamma}$, 
$x_{\gamma} > 0.75$ \cite{jeff,jet2ph}, 
hereby extracting the {\em direct} contribution to 
the cross section with just a rather small contamination from resolved 
processes. Conversely, the events with $x_{\gamma}\leq 0.75$ will represent 
the resolved part of the cross section. This procedure should therefore be 
ideal to extract $\Delta g$ on the one hand, and examine the polarized 
photon structure functions on the other.

% FIGURE 5 HERE
Fig.~5 shows the results \cite{wir} for the direct part of the cross section 
according to the above selection criteria. The contributions from the 
resolved subprocesses have been included, using the 'maximally' 
saturated set of polarized photonic parton densities. They turn out
to be non-negligible but, as expected, subdominant. More importantly,
due to the constraint $x_{\gamma}>0.75$ they are determined by the 
polarized quark, in particular the $u$-quark, distributions in the photon, 
which at large $x_{\gamma}$ are equal to their unpolarized counterparts as 
a result of the $Q^2$-evolution (see Fig.~2), rather {\em independent} of 
the hadronic input chosen. Thus the uncertainty coming from the polarized 
photon structure is minimal here and under control.
As becomes obvious from Fig.~5, the cross sections are fairly large over the 
whole range of $\bar{\eta}$ displayed and very sensitive to the shape 
{\em and} the size of $\Delta g$ with, unfortunately, not too sizeable 
asymmetries as compared to the statistical errors for ${\cal L}=100$/pb. 
A measurement of $\Delta g$ thus appears to be possible under the imposed 
conditions only if luminosities clearly exceeding $100$/pb can be reached. 
%
%FIGURE 6 HERE
Fig.~6 displays the same results, but now for the resolved 
contribution with $x_{\gamma} \leq 0.75$ for the 'maximally' saturated set 
(Figs.~6a,b) and the 'minimally' saturated one (Figs.~6c,d). As expected, 
the results depend on both the parton content of the polarized photon and 
the proton, which implies that the latter has to be known to some accuracy
to extract some information on the polarized photon structure. 
It turns out that again mostly the
polarized {\em gluon} distribution of the photon would be probed in this 
case, in particular at $\bar{\eta}>0.75$. Contributions from the 
$\Delta q^{\gamma}$ are more affected by the $x_{\gamma}$-cut; still they
amount to about $50\%$ of the cross section at $\bar{\eta}=0$.
We finally emphasize that the experimental finding of a non-vanishing 
asymmetry here would establish at least the definite existence of a resolved 
contribution to the polarized cross section. 
%
%
%\newpage
\begin{figure}[ht]
\begin{center}
\vspace*{-1.5cm}
\epsfig{file=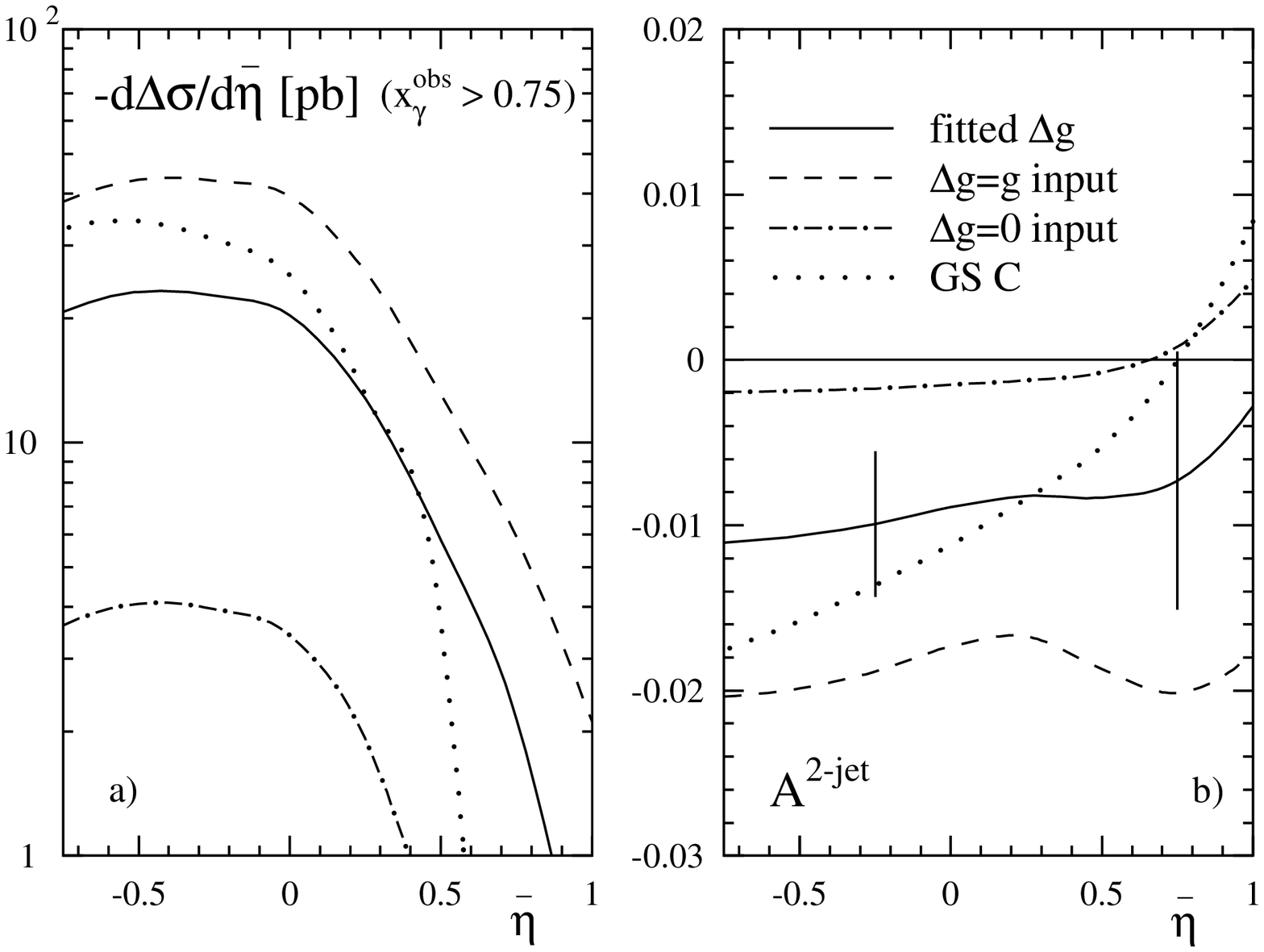,width=11.9cm}
\vspace*{-0.9cm}
\caption{\it {\bf a:} $\bar{\eta}$-dependence of the 'direct' part of the 
polarized two-jet photoproduction cross section in $ep$-collisions at HERA 
for the four different sets of polarized parton distributions of the 
proton. The experimental criterion $x_{\gamma}^{OBS}>0.75$ has been applied 
to define the 'direct' contribution (see text). The resolved contribution 
with $x_{\gamma}^{OBS}>0.75$ has been included using the 'maximally' saturated 
set of polarized photonic parton distributions. 
{\bf b:} Asymmetry corresponding to {\bf a}. The expected statistical errors 
indicated by the bars have been calculated according to (\ref{aerr}) and as 
explained in the text.}
\vspace*{-0.9cm}
%\end{center}
%\end{figure}
%
%\begin{figure}[ht]
%\begin{center}
%\vspace*{-1.5cm}
\epsfig{file=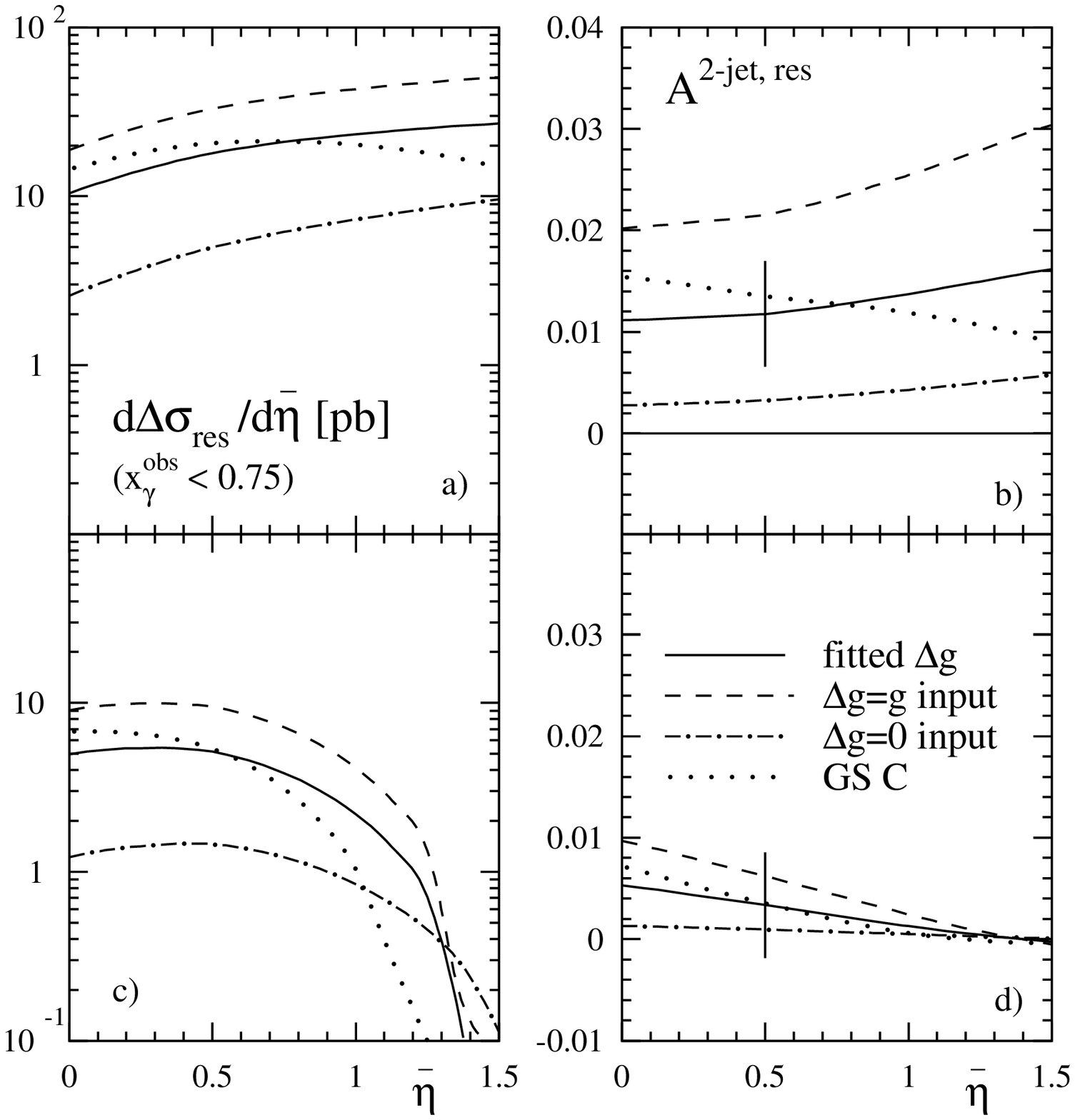,width=12.3cm}
\vspace*{-1.0cm}
\caption{\it Same as Fig.~5, but for the resolved part of the cross section, 
defined by $x_{\gamma}^{OBS}\leq 0.75$ (see text). For {\bf a,b:} the 
'maximally' saturated set of polarized photonic parton distributions has 
been used and for {\bf c,d} the 'minimally' saturated one.}
\vspace*{-0.5cm}
\vspace*{-1cm}
\end{center}
\end{figure}
\pagebreak
\section{Summary and Conclusions}
We have analyzed various photoproduction experiments in the context of a 
polarized $ep$-collider mode of HERA. All of these have already been 
successfully performed in the unpolarized case at HERA. All processes we have 
considered have in common that they get contributions from incoming gluons 
already in the lowest order and thus look promising tools to measure the 
polarized gluon distribution of the proton. In addition, they derive their 
importance from their sensitivity not only to $\Delta g$, but also to the 
completely unknown parton content of the polarized photon entering via the 
resolved contributions to the polarized cross sections. As far as a 'clear' 
determination of $\Delta g$ is concerned, this resolved piece, if 
non-negligible, might potentially act as an obstructing background, and it is
therefore crucial to assess its possible size which we have done by 
employing two very different sets for the polarized photonic parton 
distributions. Conversely, and keeping in mind that HERA has been able
to provide much new information on the unpolarized hadronic structure of the
photon, it is also conceivable that photoproduction experiments at a polarized
version of HERA could be {\em the} place to actually look for effects of the 
polarized photon structure and to prove the existence of resolved
contributions to the polarized cross sections and asymmetries. 

In the case of open-charm photoproduction we found that the resolved 
contribution is generally negligible except for the {\em total} charm cross 
section at HERA energies. Furthermore, the cross sections and their 
asymmetries are very sensitive to shape and size of $\Delta g$. We found,
however, that very high luminosities, ${\cal L}=1000$/pb, would be 
needed to measure the asymmetries with sufficient accuracy to decide 
between the various possible scenarios for $\Delta g$. Concerning 
photoproduction of jets, we find a generally much larger size of the 
resolved contribution. It turns out that the rapidity-distribution of the 
single-inclusive jet cross section separates out the direct part of the 
cross section at negative rapidities. In this region again a strong 
dependence on $\Delta g$ is found with larger cross sections than 
for the case of charm production. The corresponding asymmetries clearly
appear to be measurable here even for ${\cal L}=100$/pb. At larger 
rapidities the cross section becomes sensitive to both the parton
content of the polarized proton {\em and} photon, and an extraction 
of either of them does not seem straightforward. The situation improves 
when considering dijet production and adopting an analysis successfully
performed in the unpolarized case \cite{jeff,jet2ph} which is based on 
reconstructing the kinematics of the underlying subprocess and thus 
effectively separating direct from resolved contributions. We find that in 
this case the (experimentally defined) direct contribution should provide 
access to $\Delta g$ whereas the resolved part, if giving rise to
a non-vanishing asymmetry, would establish existence of a polarized 
parton content of the photon. Again the corresponding measurements 
would require high luminosities since the involved asymmetries are rather
small. The measurements we have proposed seem a very interesting 
challenge for a future polarized $ep$ mode of HERA.
\section*{Acknowledgements}
We are thankful to the members of the working group, in 
particular to J.\ Feltesse and M.\ Gl\"{u}ck, for helpful discussions. 
The work of M.S.\ has been supported in part by the 'Bundesministerium 
f\"{u}r Bildung, Wissenschaft, Forschung und Technologie', Bonn.

\end{document}